\documentclass[prb,twocolumn,showpacs,preprintnumbers,amsmath,amssymb]{revtex4}

\usepackage{graphicx}
\usepackage{dcolumn}
\usepackage{bm}

\begin{document}

\preprint{APS/123-QED}

\title{New high magnetic field phase of the frustrated $S=1/2$ chain compound LiCuVO$_4$ }

\author{L. E. Svistov$^{1,2}$, T. Fujita$^2$, H. Yamaguchi$^2$, S. Kimura$^2$, K. Omura$^2$, A.  Prokofiev$^4$, A. I. Smirnov$^{1,5}$,
Z. Honda$^3$ and  M. Hagiwara$^2$}
\affiliation{ $^1$P. L. Kapitza Institute for Physical Problems RAS, 119334
Moscow, Russia \\ $^2$KYOKUGEN, Osaka University, Machikaneyama 1-3, Toyanaka
560-8531, Japan \\ $^3$Graduate School of Science and Enginering, Saitama
University, 255 Shimo-okubo Saitama, Saitama 338-8570, Japan \\$^4$ Institut
f\"{u}r Festk\"{o}rperphysik Technische Universit\"{a}t Wien, A--1040 Wien,
Austria\\$^5$ Moscow Institute for Physics and Technology , 141700,
Dolgoprudny, Russia }

\date{\today}

\begin{abstract}
Magnetization of the frustrated $S=1/2$  chain compound LiCuVO$_4$, focusing on high magnetic field phases, is
reported. Besides a spin-flop transition and the transition from a planar spiral to a spin modulated structure
observed recently, an additional transition  was observed just below the saturation field. This newly observed
magnetic phase is considered as a spin nematic phase, which was predicted theoretically but was not observed
experimentally. The critical fields of this phase and its dM/dH curve are in good agreement with calculations
performed in a microscopic model (M. E. Zhitomirsky and H. Tsunetsugu, preprint, arXiv:1003.4096v2).

\end{abstract}

\maketitle

Unconventional magnetic orders and phases in frustrated quantum spin chains are attractive issues, because they
appear under a fine balance of the exchange interactions and are sometimes caused by much weaker interactions or
fluctuations.\cite{Chubukov_1991,Kolezhuk_2000,Kolezhuk_2005,Dmitriev_2008}

A  kind of frustration in quasi one--dimensional (1D) magnets is provided by competing interactions when the
intra--chain nearest neighbor (NN) exchange is ferromagnetic and the next--nearest neighbor (NNN) exchange is
antiferromagnetic.  This type of exchange results in the formation of bound magnon pairs in a high magnetic
field near the saturation point.\cite{Chubukov_1991, Kuzian_2007,Ueda_2009}  The condensation of bound magnon
pairs could provide a nematic spin ordering.  Numerical investigations of frustrated chain magnets with
different models~\cite{Hikihara_2008, Sudan_2009, Heidrich_2009} have, indeed, predicted a nematic ordering of
transverse spin components at zero temperature in a magnetic field range close to the saturation field. The spin
nematic phase was introduced phenomenologically~\cite{Andreev_1984} as a magnetic state with zero ordered spin
components and nonzero spin correlations emerging via a second order phase transition. This spin ordering is
analogous to the well known ordering of molecules in nematic phases of liquid crystals.\cite{Gennes_1974}
Besides the frustrated spin $S=1/2$ chains,  $S\geqslant 1$ systems with biquadratic exchange were proposed as
potential nematics.\cite{Blume_1969} Nevertheless, the real spin nematics have not been found during a long
period.

 Recently a new analytical approach for
the description of the high field magnetic phase of frustrated chains with a spin-nematic ordering was
proposed.\cite{Zhitomirsky_2010} The parameters of this curious phase were calculated from the representation of
the condensate wave function in terms of the coherent bosonic states. A $S=1/2$ quasi 1D magnet with the
competing interactions of the described type, LiCuVO$_4$, was analyzed as a potential spin nematic. For
LiCuVO$_4$, the nematic phase was predicted to exist in the field interval of about 3 T below the saturation
field. This work stimulated our high field magnetization measurements on LiCuVO$_4$ single crystals.

\begin{figure}
\includegraphics[width=80 mm,angle=0,clip]{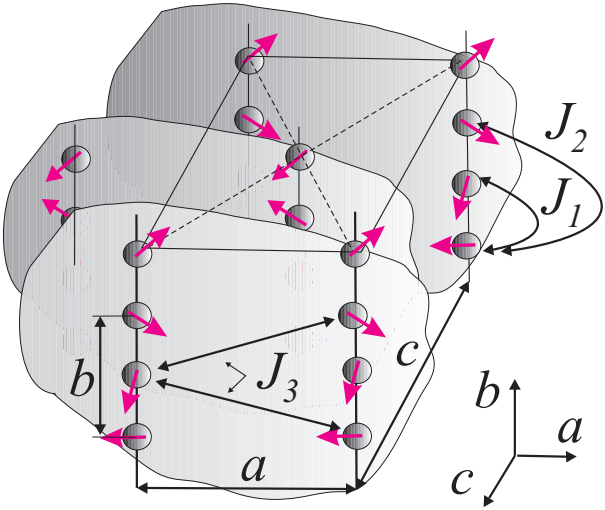}
\caption{(Colour online) Scheme of  Cu$^{2+}$ moments in the crystal structure of LiCuVO$_4$. Copper ions are
marked by circles. Arrows constitute a helical spin structure at $H=0$ below $T_{\rm N}$
(Ref.~\onlinecite{Gibson_2004}). $J_1, J_2$, and $J_3$ are main exchange integrals~\cite{Enderle_2005}. The
exchange bonds $J_1, J_2$, $J_3$ lie in shadowed planes, which are parallel to the ab-planes.} \label{Fig_1}
\end{figure}

LiCuVO$_4$ crystallizes in an inverse spinel structure $AB_2$O$_4$ with an
orthorhombic distortion. Copper ions share the octahedral sites being arranged
in chains directed along the $b$--axis. These chains are separated by the
nonmagnetic ions of Li, V, O. The elementary cell contains four ions
Cu$^{2+}$($S=1/2$) (see Fig.~\ref{Fig_1}). Neutron--diffraction
experiments~\cite{Gibson_2004} indicate that in the ordered phase at $T <
T_N=2.3$ K and $H=0$ an incommensurate planar spiral spin structure occurs with
the wave vector $k_{\rm ic}$ = $(0.468) \cdot 2\pi/b$ directed along the
$b$--axis. The Curie-Weiss temperature is $\Theta_{CW}=-15$~K. The ordered
magnetic moments of Cu$^{2+}$ ions lie within the $ab$--planes and amount to
0.31$\mu_B$ at $T$=1.6 K. The strong reduction of the ordered component and a
small value of  $T_N / \Theta_{CW}$ indicate a significant influence of quantum
fluctuations and frustration on the ground state.

\begin{figure}
\includegraphics[width=80 mm,angle=0,clip]{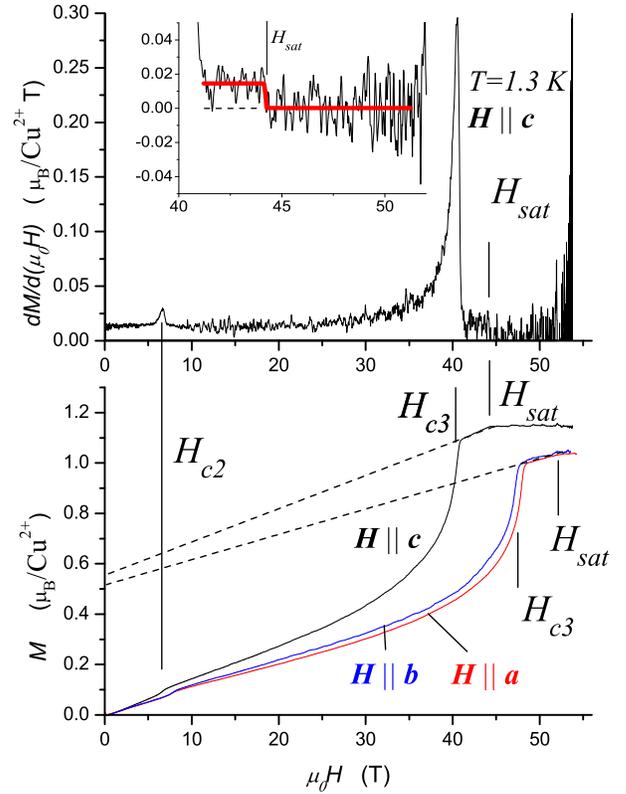}
\caption{(Color online.)Upper panel: $dM/dH$ curve measured with a pulse field technique. Insert: $dM/dH$ near
$H_{sat}$. The solid line presents the fit by a step-like function. The dashed line marks zero level.  Lower panel:
$M(H)$ curves at $T=$1.3 K obtained by integration  for $H\parallel a,b, c$. The solid lines correspond to the magnetization for field descending process. The straight dashed lines present linear fits in the range
$H_{c3}<H<H_{sat}$ and extrapolations of these fits to zero field.}   \label{Fig_2}
\end{figure}

\begin{figure}
\includegraphics[width=80 mm,angle=0,clip]{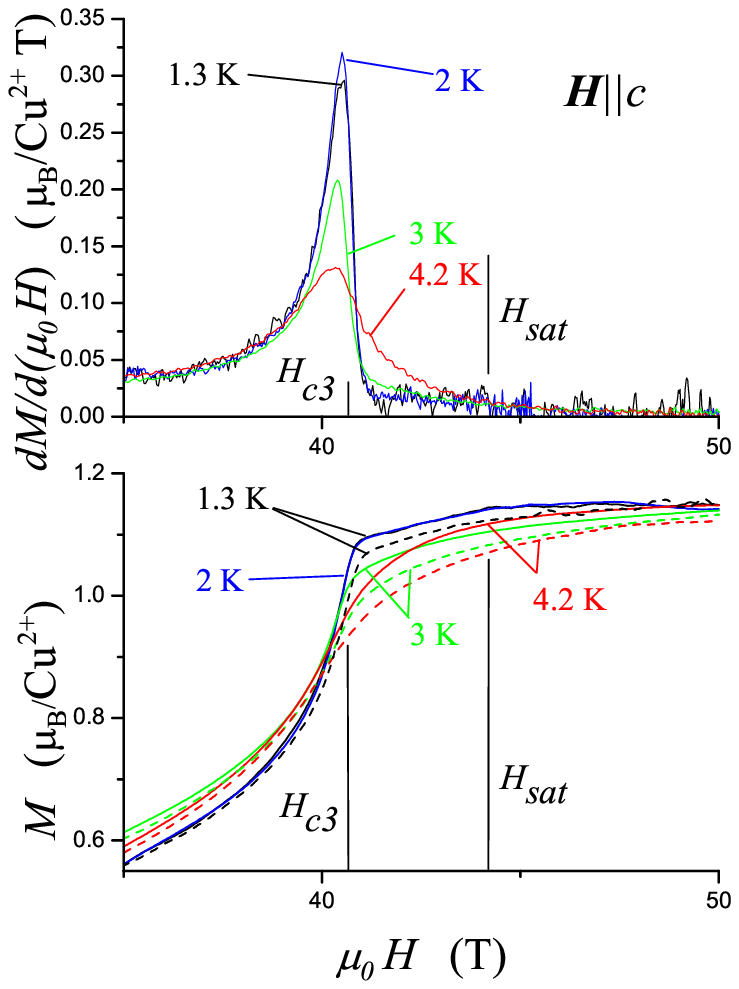}
\caption{(Colour online.) $dM/dH$, $M(H)$ curves measured with a pulse field
technique for $H$$\parallel$$c$. The dashed and solid lines correspond to the
magnetization curves for field ascending and descending processes,
respectively.} \label{fig_3}
\end{figure}

Inelastic neutron--scattering experiments~\cite{Enderle_2005} confirmed that the exchange interactions $J_{1,2}$
within chains are dominant, and the NN and NNN intrachain exchange interactions are ferromagnetic and
antiferromagnetic, respectively. It was shown that the incommensurate structure is due to these competing
interactions. In addition, it was found that there is a ferromagnetic exchange between neighboring chains along
the $a$--axis ($J_3$), and an antiferromagnetic exchange link between different $ab$--planes ($J_4$), which is
almost five times smaller than $J_3$.

In a magnetic field $H$, consecutive magnetic phase transitions were
observed.~\cite{Enderle_2005, Buettgen_2007, Banks_2007, Buettgen_2010} Below
12~T, the magnetic phases were studied also by means of ESR and NMR
techniques.~\cite{Buettgen_2007} For $H$  within the $ab$-plane, the first
transition takes place at $\mu_0H_{c1}\approx 2.5$~T, being interpreted as a
spin--flop reorientation of the spin spiral plane perpendicular to
$H$.~\cite{Buettgen_2007}  At $\mu_0H_{c2}\approx 7$~T, there is a transition
to a phase with the collinear spin-modulated structure. The spins in this phase
are directed along the magnetic field. This ordering  occurs only within
separate $ab$-planes, and adjacent planes are not
correlated.~\cite{Buettgen_2007, Buettgen_2010} A week anisotropy of $H_{c2}$
reveals an exchange nature of this transition. A possible mechanism of this
transition was proposed in Refs.~\onlinecite{Hikihara_2008} and
\onlinecite{Heidrich_2009}. Further, the saturation of the magnetization was
ascribed to $dM/dH$  anomalies observed at $\mu_0H_ \approx 40$~T and 45~T for
the orientations $H\parallel c$ and $H\parallel a$,
respectively.~\cite{Enderle_2005, Banks_2007} However, the true saturation,
i.e. the transition to constant M, was not achieved there.

Single crystals of LiCuVO$_4$ of several cubic millimeters were prepared as
described in Ref.~\onlinecite{Prokofiev_2005}. The composition of samples is
characterized by the relations: Li/V=0.96$\pm$0.05, Li/Cu=0.95$\pm$0.04,
Cu/V=0.99$\pm$0.01. Thus, the average composition is Li$_{0.97}$CuVO$_4$, which
means a low concentration of vacancies in the Li-sublattice. This type of
crystals has the unit cell parameter $c$=8.745 \AA. The analysis of the quality
of samples is discussed in Ref.~\onlinecite{Prokofiev_2004}. The samples used
in the present work are from the butch 1 of Ref.~\onlinecite{Prokofiev_2004}.
Magnetization curves in magnetic fields of up to 7~T were measured with a
commercial SQUID magnetometer (Quantum Design MPMS-XL7). High field
magnetization data were taken in magnetic fields of up to 55~T using a pulsed
magnet at KYOKUGEN in Osaka University.~\cite{Hagiwara_2006}

Experimental $dM/dH$  for $H \parallel c$ curve and its field integral $M(H)$ for $H \parallel a,b,c$ are shown in
Fig.~2. Each $dM/dH$ curve demonstrates two sharp peaks at the fields $H_{c2}$, $H_{c3}$  and a step at
$H_{sat}$. A strong inflection is seen on the $M(H)$ curve below the field $H_{c3}$. The lowest transition,
marked as $H_{c2}$, occurs at $H\approx7$ T for all three field directions. This anomaly corresponds to the
transition from a spiral to a collinear spin modulated phase described above. The field derivative of
magnetization above $H_{c2}$ is a bit smaller than that in the low field phase. This difference equals 5$\pm$1\%
for $H\parallel c$, $3.5\pm 1$\% for $H\parallel a$ and 2.5$\pm$ 1\% for $H\parallel b$. At this critical field,
a jump of magnetization for 14$\pm$2\% occurs.

Our intriguing finding in the magnetization curve is a "{\it fine structure}" near the saturation  process,
which includes anomalies at $H_{c3}$ and $H_{sat}$.  The field $H_{c3}$ is marked by the maximum of  $dM/dH$.
Right under $H_{c3}$ (Fig. 2), the $dM/dH$ increases abruptly more than by a factor of twenty. The magnetic
moment at $H_{c3}$ reaches  95\% of the saturation value. Above $H_{c3}$, the value of $dM/dH$ drops quickly to
that in the middle field range. Then, it goes down to zero at the saturation field $H_{sat}$. The value of
$H_{sat}$ is determined by the fit of $dM/dH$ curve in the interval above $H_{c3}$ by a Fermi-like step
function: $dM/dH=\chi/(exp((H-H_{sat})/\Delta H)+1)$ where $H_{sat}$, the derivative  $dM/dH=\chi$, and the width of
the saturation transition $\Delta H$ are fitting parameters. The obtained value of $H_{sat}$ does not depend on
the boundaries of the fitting range within 1\%. The value of $\chi$  is obtained with the accuracy of 20\%,
$\Delta H$ is less then $0.01$ T. The magnetic moment in the saturated phase for $H$$\parallel$$c$ is $M=1.15\pm
0.05~\mu_B$/Cu$^{2+}$, consistent with the expected $1.15~\mu_B$ for $g_c$=2.3. The fields $H_{c3}$ and
$H_{sat}$ are well separated by an interval of 3.9 $\pm 0.1$~T for $H\parallel c$, 4.9$\pm 0.2$~T for
$H\parallel a,b$.

The temperature evolution of the $M(H)$ curve for $H\parallel c$ is shown in
Fig.~3. The small kink of $M(H)$ at $H_{sat}$ is distinguishable below 2 K.
Above 2 K, the phase transition at $H_{c3}$ is broadened and the anomaly at
$H_{sat}$ is masked. At 1.3 K, the hysteresis is practically negligible, while
at higher temperatures it becomes larger and is probably caused by the
magnetocaloric effect.

 The results of low field measurements are depicted in Fig.~4.
The $dM/dH$ curves for $H\parallel a$ and $H\parallel b$ have peaks at
$H_{c1}$=2.5 T and 3.1 T, respectively, corresponding  to the spiral-plane flop
transition. The critical fields are listed in Table~1. The proposed magnetic
phases at $T < 2$~K are presented in the lower panel of Fig.~4. For $H\parallel
c$, the low field transition is absent, because the spiral-plane flop does not
take place in this case. The values of $H_{c1}$ for $a$ and $b$ directions are
consistent with those obtained from ESR measurements.~\cite{Buettgen_2007} The
$H_{c2}$ for $H \parallel c$ is close to that observed in the $dM/dH$ curve in
Ref.~\onlinecite{Banks_2007}. The field $H_{c3}$  corresponds well to the peak
of $dM/dH$ observed in Ref.~\onlinecite{Enderle_2005}, where this anomaly was
ascribed to the saturation field. Nevertheless, the right wing of this $dM/dH$
peak at the high field range of their work appeared much more diffuse than that
in the present observation. The interval where the $dM/dH$ decreases twice is
for this experiment about 2 T while for our observation this interval is 0.25
T. A large width of the right wing of the $dM/dH$ peak masks the details of
saturation in the experiments of Refs.~\onlinecite{Banks_2007} and
\onlinecite{Enderle_2005}. Our measurements make clear that the field at the
large peak of $dM/dH$ is not the saturation field $H_{sat}$ as supposed before,
but marks a transition into a new high field phase.

For the analysis, we use the model Hamiltonian:
\begin{eqnarray}\label{eq:1}
\hat{\cal{H}} & = & \sum[J_1(\textbf{\textit S}_{i,j}\cdot \textbf{\textit S}_{i,j+1})+\textbf{\textit S}_{i,j}
\cdot \textbf{\textit J}_{1}^{a} \cdot \textbf{\textit S}_{i,j+1}] \\ \nonumber & + &
\sum[J_2(\textbf{\textit S}_{i,j}\cdot \textbf{\textit S}_{i,j+2})+\textbf{\textit S}_{i,j} \cdot \textbf{\textit J}_{2}^{a} \cdot
\textbf{\textit S}_{i,j+2}] \\ & + & \sum[J_3(\textbf{\textit S}_{i,j}\cdot
\textbf{\textit S}_{i+1,j+1})+J_3(\textbf{\textit S}_{i+1,j+1}\cdot \textbf{\textit S}_{i,j})] \nonumber \\ & + & \sum
\mu_{B}\textbf{\textit H}\cdot \textbf{\textit g} \cdot \textbf{\textit S}_{i,j},\nonumber
\end{eqnarray}
where $g_{aa}=2.070$, $g_{bb}=2.095$, and $g_{cc}=2.313$ are the diagonal components of $g$-tensor, $J_1/k_{\rm
B}=-18.5$~K, $J_2/k_{\rm B}=44$~K, $J_3/k_{\rm B}=-4.3$~K and $J_{1aa}^a/k_{\rm B}=0.16$~K,  $J_{1bb}^a/k_{\rm
B}=-0.02$~K, and $J_{1cc}^a/k_{\rm B}=-1.75$~K, $J_{2cc}^a/k_{\rm B}=-0.2$~K are the diagonal components of
$\textbf{\textit J}_1^a$ and $\textbf{\textit J}_2^a$ -tensors. The remaining components of tensors are proposed
to be small. The summation is over exchange bonds marked in Fig. 1. The components of $\textbf{\textit g}$ and
$\textbf{\textit J}_{1}^a$ tensors are obtained from ESR experiments~\cite{Krug_2002}, $J_{1,2,3}$ -- from the
neutron scattering study~\cite{Enderle_2005} and $J_{2cc}$ -- from the gap of the low--frequency
antiferromagnetic resonance mode~\cite{Buettgen_2007}. The saturation fields $H^{a,b,c}_{sat}$ may be evaluated
roughly in the classical 1D approximation as $H^{i}_{sat}$=4$J_2S(1+J_1/4J_2)^2/g_{i}\mu_B$ ($i$=$a, b,$ and $
c$)~\cite{Nagamiya_1962, Zhitomirsky_2010}: $\mu_0 H_{sat}^a$=50 T, $\mu_0 H_{sat}^b$=50 T, and $\mu_0
H_{sat}^c$=45 T.  These values are in good agreement with our experiment.

\begin{figure}
\includegraphics[width=80 mm,angle=0,clip]{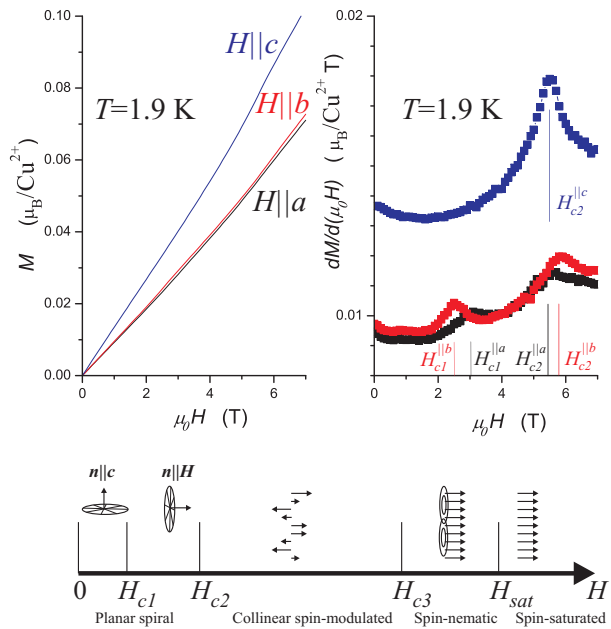}
\caption{ (Colour online) Upper panel: the low field magnetization curves
$M(H)$ (left) and their field derivatives (right) for three field directions
and at $T$=1.9~K. Lower panel: the scheme of expected magnetic phases in
LiCuVO$_4$ at $T < 2$~K. } \label{fig_4}
\end{figure}

\begin{small}

\begin{table}[h]
\caption{}

\begin{tabular}{|c|c|c|c|c|}
\hline  & $H_{c1}$ (1.9~K)  & $H_{c2}$ (1.3~K)  & $H_{c3}$ (1.3~K) & $H_{sat}$ (1.3~K)
\\ \hline
  $H$$\parallel$$a$ & 2.5~T  & 7.6$\pm$0.2~T   & 47.9$\pm$0.2~T  & 52.8$\pm$0.3~T \\
  $H$$\parallel$$b$& 3.1~T  & 7.6$\pm$0.2~T  & 47.2$\pm$0.2~T  & 52.1$\pm$0.3~T \\
  $H$$\parallel$$c$& - & 6.7$\pm$0.2~T  & 40.5$\pm$0.2~T  & 44.4$\pm$0.3~T  \\  \hline
  Theory$^{13}$ & - & - & 44.5~T  & 47.1~T\\
  {\footnotesize $(T=0)$}& & & &\\
   \hline
\end{tabular}

\end{table}
\end{small}

Our observations clearly indicate that the field $H_{c3}$ is different from the saturation field. The field and
magnetization intervals, corresponding to the newly found magnetic phase, can be compared with theoretical
calculations performed in different models for a nematic phase. A numerical investigation of a frustrated chain
was carried out~\cite{Heidrich_2009} using a model with anisotropic exchange interactions $J_{1cc}^a$,
$J_{2cc}^a$, at the condition $J_{1cc}^a$/$J_1$=$J_{2cc}^a$/$J_2$.
 Three phases were predicted for a system with exchange parameters $J_{1}$, $J_{2}$, and $J_{1cc}^a$
 of LiCuVO$_4$. In low fields, there is  a quadrupolar spin density wave phase, having the same nearest spin
 correlation as the spin modulated phase observed in the experiment. A
quadrupolar spin nematic phase should occur in a range below $H_{sat}$. Then, a
saturated phase appears above $H_{sat}$. This work predicts a nematic phase in
the interval of magnetization between $M_{sat}-\Delta M$ and $M_{sat}$. The
value of $\Delta M$ is predicted to be 20-30 \% of $M_{sat}$. The experimental
value of $\Delta M$ in LiCuVO$_4$, however, is 5\% of $M_{sat}$. In low fields,
a  planar spiral is observed for LiCuVO$_4$ instead of a spin density wave
predicted by the model. Note that the experimental $M(H)$ curves differ
essentially from those obtained numerically for a 1D
model~\cite{Heidrich_2009}. For 1D model, the fluctuation divergence of $dM/dH$
takes place at $H_{sat}$, whereas in the experiment the divergence is found at
$H_{c3} <H_{sat}$. Probably, this difference is caused by the 2D character of
exchange links in LiCuVO$_4$.

A more adequate 2D model with exchange interactions $J_1$, $J_2$, $J_3$ was
considered analytically in Ref.~\onlinecite{Zhitomirsky_2010}. A phase
transition from a spiral umbrella-like phase to a spin-nematic one was
predicted at a critical field $H_{c}$. The field range of the nematic phase
should reach $H_{sat}$. For exchange parameters of LiCuVO$_4$, the saturation
field is predicted to be 47.1 T, and the width of field range of nematic phase
$H_{sat}-H_{c}$=2.6 T, which is not far from the experimental value of
$H_{sat}$-$H_{c3}$ (3.9 -- 4.9~T). This theory presumes an isotropic $g$-factor
of 2.0, which differs from the real values of $g_{a,b,c}$. Besides, the
exchange parameters $J_{1,2,3}$ are known to have the accuracy of 12, 1.5 and
70\%,
 correspondingly. This may be a reason of a considerable uncertainty in the estimation of $H_c$.  Moreover,
we compare the value of $dM/dH$ in the high field phase with the theoretical prediction for the spin nematic
phase. According to Ref.~\onlinecite{Zhitomirsky_2010},  $dM/dH$ in the nematic state is approximately a half of
the slope of the classical magnetization curve, which is $M_{sat}/H_{sat}$. As one can see in Fig.2, the value
of dM/dH in the high field phase corresponds well to $\frac{1}{2}M_{sat}/H_{sat}$ indeed.

In conclusion, a new magnetic phase near the saturation field is observed in a quasi-1D $S=1/2$ magnet with
competing ferromagnetic NN and antiferromagnetic NNN exchange interactions LiCuVO$_4$. By comparison with
theoretical studies of Refs.~\onlinecite{Chubukov_1991, Hikihara_2008, Sudan_2009,
Heidrich_2009,Zhitomirsky_2010}, this new phase must be a spin nematic one. This proposition is supported by
reasonable agreement between the critical fields and dM/dH calculated analytically~\cite{Zhitomirsky_2010} in
the 2D model and experimental ones.

 We thank M. Zhitomirsky and T. Momoi for useful discussions. This work was carried out
 under the Visiting Researcher Program of KYOKUGEN and
partly supported by Grants-in-Aid for Scientific Research (No.20340089), the Global COE Program (Core Research
and Engineering of Advanced Materials-Interdisciplinary Education Center for Materials Science) (No. G10) from
the MEXT, Japan, by the Grants 09-02-12341, 10-02-01105-a of the Russian Foundation for Basic Research, and
Program of Russian Scientific Schools.


\begin{thebibliography}{25}
\bibitem{Chubukov_1991} A. V. Chubukov, Phys. Rev. B {\bf{44}}, 4693
(1991).
\bibitem{Kolezhuk_2000} A. K. Kolezhuk, Phys. Rev. B {\bf{62}}, 6057(R)
(2000).
\bibitem{Kolezhuk_2005} A. K. Kolezhuk, T. Vekua, Phys. Rev. B {\bf{72}}, 094424
(2005).
\bibitem{Dmitriev_2008} D. V. Dmitriev, V. Yu. Krivnov, Phys. Rev. B {\bf{77}}, 024401
(2008).
\bibitem{Kuzian_2007} R. O. Kuzian, S.-L. Drechsler, Phys. Rev. B {\bf{75}}, 024401
(2007).
\bibitem{Ueda_2009} H. T. Ueda, K. Totsuka, Phys. Rev. B {\bf{80}}, 014417
(2009).
\bibitem{Hikihara_2008} T. Hikihara, L. Kecke, T. Momoi, A. Furusaki, Phys. Rev. B {\bf{78}}, 144404
(2008).
\bibitem{Sudan_2009} J. Sudan, A. L\"{u}scher,  A. M. L\"{a}uchli, Phys. Rev. B {\bf 80},
140402(R) (2009).
\bibitem{Heidrich_2009} F. Heidrich--Meisner, I. P. McCulloch,  A. K. Kolezhuk,
Phys. Rev. B {\bf 80}, 144417 (2009).
\bibitem{Andreev_1984} A. F. Andreev, I. A. Grishchuk, Sov. Phys. JETP {\bf 60}, 267 (1984).
\bibitem{Gennes_1974} P. G. de Gennes, "Liquid crystals", Clarendon Press, Oxford (1974).
\bibitem{Blume_1969}M. Blume, Y. Y. Hsieh, J. Appl. Phys. {\bf 40}, 1249 (1969).
\bibitem{Zhitomirsky_2010} M. E. Zhitomirsky, H. Tsunetsugu, arXiv:1003.4096v1
[cond-mat.str-el] 22 Mar 2010.
\bibitem{Gibson_2004} B. J. Gibson, R. K. Kremer, A. V. Prokofiev, W. Assmus, G. J. McIntyre, Physica {\bf B} {\bf 350}, e253 (2004).
\bibitem{Enderle_2005} M. Enderle, C. Mukherjee, B. F\"ak, R. K. Kremer, J.-M. Broto, H. Rosner, S.-L. Drechsler,
J. Richter, J. Malek, A. Prokofiev, W. Assmus, S. Pujol, J.-L. Raggazzoni, H.
Rakoto, M. Rheinst\"adter, H. M. R{\o}nnow, Europhys. Lett. {\bf 70}, 237
(2005).
\bibitem{Buettgen_2007} N. B\"{u}ttgen, H.-A. Krug von Nidda, L. E. Svistov, L. A. Prozorova, A. Prokofiev,
W. Assmus, Phys. Rev. B {\bf{76}}, 014440 (2007).
\bibitem{Banks_2007} M. G. Banks, F. Heidrich-Meisner, A. Honnecker,
H. Rakoto, J.-M. Broto, R. K. Kremer, J. Phys.: Condens. Matter {\bf 19},
145227 (2007).
\bibitem{Buettgen_2010} N. B\"{u}ttgen, W. Kraetschmer, L. E. Svistov, L. A. Prozorova, A. Prokofiev,
 Phys. Rev. B {\bf{81}}, 052403 (2010).
\bibitem{Prokofiev_2005} A. V. Prokofiev, I. G. Vasilyeva, W. Assmus, J. Crystal Growth {\bf{275}}, e2009 (2005).
\bibitem{Prokofiev_2004} A. V. Prokofiev, I. G. Vasilyeva, V. N. Ikorskii, V. V. Malakhov, I. P. Asanov, W. Assmus, Sol. State Chem. {\bf{177}}, 3131 (2004).
\bibitem{Hagiwara_2006} M. Hagiwara, S. Kimura, H. Yashiro, S. Yoshii, K. Kindo,  J. Phys. Conf. Ser. {\bf 51}, 647 (2006).
\bibitem{Krug_2002} H.-A. Krug von Nidda, L. E. Svistov, M. V. Eremin, R. M. Eremina,
A. Loidl, V. Kataev, A. Validov, A. Prokofiev, W. Assmus, Phys. Rev. B {\bf65},
134445 (2002).
\bibitem{Furukawa_2008} S. Furukawa, M. Sato, Y. Saiga, S. Onoda, J. Phys. Soc. Jpn. {\bf 77}, 123712 (2008).
\bibitem{Nagamiya_1962} T. Nagamiya, K. Nagata, Y. Kitano, Prog. Theor. Phys. {\bf{27}}, 1253 (1962).









\end{thebibliography}
\end{document}